\documentclass[article,twocolumn, prl]{revtex4}
\usepackage{epsfig}
\usepackage{amsmath}
\usepackage{epsfig}
\usepackage{mathrsfs}
\usepackage{comment}
\usepackage{sidecap}
\begin{document}

\newcommand{\bdm}{\begin{displaymath}}
\newcommand{\edm}{\end{displaymath}}

\newcommand{\be}{\begin{equation}}
\newcommand{\ee}{\end{equation}}

\newcommand{\bi}{\begin{itemize}}
\newcommand{\ei}{\end{itemize}}

\title{Mechanical cat states in graphene resonators}
\author{A. Voje}
\author{J. M. Kinaret} 
\author{A. Isacsson$^*$}
\affiliation{Department of Applied Physics, Chalmers University of
  Technology,
  SE-412 96 G{\"o}teborg Sweden.\\
  $^*$Corresponding author: andreas.isacsson@chalmers.se}
\date{Version: \today}
\begin{abstract}
We study the quantum dynamics of a symmetric nanomechanical graphene
resonator with degenerate flexural modes. Applying voltage pulses to
two back gates, flexural vibrations of the membrane can be selectively
actuated and manipulated. For graphene, nonlinear response becomes
important already for amplitudes comparable to the magnitude of zero
point fluctuations. We show, using analytical and numerical methods,
that this allows for creation of cat-like superpositions of coherent
states as well as superpositions of coherent cat-like non-product states.
\end{abstract}

\maketitle 
Coherent superposition of states are characteristic traits of quantum
mechanics. These phenomena have already been realized in many-particle
contexts such as trapped ultra-cold atoms~\cite{Monroe_1996},
superconductors~\cite{Friedman_2000} and photonic
systems~\cite{Haroche_2008}. A current challenge is to observe these
effects for collective degrees of freedom in a macroscopic context in,
e.g., mechanical resonators~\cite{Schwab_2005}.

Recent advances in cooling mechanical resonators and sensitive
displacement detection have allowed reaching the motional ground state
and observing zero point fluctuations of center of
mass~\cite{Cleland_2010, Lehnert_2011}. Active manipulation and
characterization of the quantum state of these systems, as already
achieved with photons~\cite{Cleland_2009}, seem to be within
reach. For a mechanical system, a desirable state to generate is a
'cat' state. This is a coherent superposition of two minimum
uncertainty wave packets separated by more than their individual
quantum fluctuations.

For the harmonic oscillator a minimum uncertainty wave packet is a
coherent state $\left|\alpha\right> =\exp[\alpha
  a^{\dag}-\alpha^{\ast}a]\vert 0 \rangle$ generated by displacing the
oscillator ground state~\cite{Gardiner_book}. As shown by Yurke and
Stoler~\cite{Yurke_Stoler}, for a nonlinear oscillator
$H=\hbar\omega_0\hat{n}+\hbar\Omega\hat{n}^2$ an initial coherent
state $\left|\alpha\right>$ will after a time $t=\pi/(2\Omega)$ evolve
into the cat state
$(1/\sqrt{2})\left[e^{-i\pi/4}\left|\alpha\right>+e^{i\pi/4}\left|-\alpha\right>\right]$
with the maximum spatial separation $\Delta=2|\alpha|$.
 
Nanoelectromechanical resonators are typically intrinsically
nonlinear~\cite{Cleland_book}. The amplitudes needed to observe
nonlinear effects are often orders of magnitude larger than the
quantum zero point fluctuations $x_0=\sqrt{\hbar/m\omega_0}$. A cat
state obtained due to this nonlinearity would have a
separation $\Delta\gg x_0$. As
the decoherence rate scales as $(\Delta/x_0)^2$~\cite{Holmes_1986,
  Kartner_1993} this has, until now, been unfeasible.
Instead, coupling to auxillary quantum systems
has been proposed to engineer the
nonlinearity~\cite{Jacobs_2007,Milburn_2009}.

We show how the intrinsic nonlinearity
in a graphene membrane resonator can be used to prepare cat states
by applying voltage pulses to local backgates. The reason for using
graphene is the ultra-low mass of the graphene sheet which leads to a
large $x_0$, and an onset of nonlinear response at small
amplitudes~\cite{Atalaya_2008, Bachtold_2011}. This implies that cat
states with moderate ratios $(\Delta/x_0)$ can be constructed without
the need for engineered auxillary quantum systems or feedback
loops. Another feature of two-dimensional mechanical resonators is
that they can be designed to have degenerate flexural modes. Coupling
between these modes can be controlled by external
perturbations such as gate electrodes, which can be utilized for state
manipulation.

\begin{figure}[t]
\begin{center}
\epsfig{file=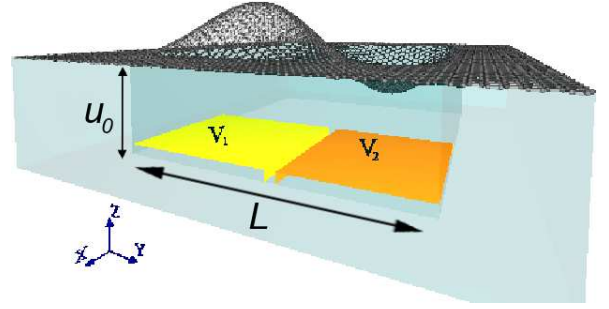,width=0.9\linewidth} 
\end{center}
\caption{(Color online) Cross section of a square graphene membrane
  resonator. A fully clamped graphene membrane with side $L$ is
  suspended a distance $u_0$ above the substrate. Below, covering two
  adjacent quadrants beneath the membrane, are local backgates with
  time dependent voltage biases $V_{1,2}(t)$. By applying pulses to
  the local gates, cat-like superpositions of flexural mode states, as
  well as superpositions of coherent cat-like non-product states, can
  be generated.
  \label{fig:system}}
\end{figure}

For concreteness we consider a square graphene membrane with mass
density $\rho_0=7.6\cdot 10^{-7}$ kg/m$^2$ and side length $L$. The
sheet is suspended in the $xy$-plane at a distance $u_0$ above two
local backgates (see Fig.~\ref{fig:system}). They cover adjacent
quadrants below the membrane and are have voltages $V_1(t)$ and
$V_2(t)$. The Hamiltonian density of the system can be divided into
two parts: $\mathscr{H} = \mathscr{H}_0({\bf x}) + \mathscr{H}_G({\bf
  x},V_1,V_2)$, where ${\bf x}\equiv (x,y)$. Here $\mathscr{H}_0({\bf
  x})$ gives the intrinsic mechanics of the membrane while the
coupling to the gates is described in $\mathscr{H}_G({\bf
  x},V_1,V_2)$.  To a first approximation one
finds~\cite{Atalaya_2008,Daniel_book}
\begin{eqnarray}
\mathscr{H}_0 &=& \frac{\pi_0^2}{2\rho_0} +
\frac{T_0}{2}\big|\nabla u\big|^2+
\frac{T_1}{4}\big|\nabla u\big|^4
\label{Hdensity0} 
\end{eqnarray}
where $u({\bf x},t)$ is the out of plane displacement and $\pi_0$ its
conjugate momentum density. The built-in tension is $T_0$ and the
stretching-induced tension is determined by $T_1 = (\mu + \lambda/2)$ where
the Lam\'e-parameters of graphene are $\mu\approx 3\lambda \approx 9$
eV/{\AA}$^2$~\cite{Yakobsson_2001}. We model $\mathscr{H}_G$ in the
local approximation as $\mathscr{H}_G ={\epsilon_0 V({\bf
    x},t)^2}/[{u_0 +u({\bf x})]}$, where $\epsilon_0 = 8.854\cdot
10^{-12}$~F/m and $V({\bf x},t)$ is the local gate potential.
Expanding $u$ and $\pi_0$  in mode functions as
\begin{equation}\label{modeExpansion}
u = \sum_{\bf{k}}q_{\bf{k}}(t)\varphi_{\bf{k}}({\bf{x}}),\,\,\,
\pi_0=\displaystyle{\frac{1}{L^2}}\sum_{\bf{k}}p_{\bf{k}}(t)\varphi_{\bf{k}}({\bf{x}}),\nonumber
\end{equation}
using $\varphi_{\bf{k}}= 2\sin(k_x x)\sin(k_y y)$ and ${\bf
  k}=(k_x,k_y)=(n,m)\pi/L$ for $n,m=1,2,...$,
gives $H=H_0+H_G$ with
\begin{eqnarray}
&&H_0 = \sum_{\bf{k}}\frac{1}{2M}\Big( p_{\bf{k}}^2
+M^2\omega_{\bf{k}}^2 q_{\bf{k}}^2\Big) +\frac{1}{2}\sum_{\bf{kk^{'}}}
F_{\bf{kk^{'}}} (q_{\bf{k}} q_{\bf{k^{'}}})^2,\nonumber\\
&&H_G = K_0 + \sum_{\bf{k}} K_{\bf{k}} q_{\bf{k}} + \frac{1}{2}\sum_{\bf{kk^{'}}} K_{\bf{kk^{'}}}
q_{\bf{k}} q_{\bf{k^{'}}}.\nonumber
\end{eqnarray}
Here $M = \rho_0 L^2$ and $\omega_{\bf k}
=({T_0}/{\rho_0})^{\frac{1}{2}}\bf |{k}|$. The coefficients
$F_{\bf{kk^{'}}}$ and $K_{\bf{kk^{'}}}$
will be discussed below.

We restrict attention to the two lowest degenerate modes with
${\bf{k}}_1= (\pi/L,2\pi/L)$, and ${\bf{k}}_2=(2\pi/L,\pi/L)$ and
their frequencies $\omega_{1,2}= (\pi/L)\sqrt{5T_0/\rho_0}$. We label
these modes ${\bf1}$ and ${\bf2}$. Quantizing, by imposing the
commutation relation
$[\hat{q}_{\bf{k}},\hat{p}_{\bf{k'}}]=i\hbar\delta_{\bf k,\bf k'}$,
yields
\begin{eqnarray}
&&H=\frac{ \hat{p}_{1}^2 + \hat{p}_{2}^2 }{2M} +
\frac{M\omega_1^2}{2}(\hat{q}_{1}^2 +\hat{q}_{2}^2)
+ D(\hat{q}_{1}^4 +\hat{q}_{2}^4)+ F\hat{q}_{1}^2\hat{q}_{2}^2 \nonumber \\
&&+ K_1\hat{q}_{1} +K_2 \hat{q}_{2} +
K_{12}\hat{q}_{1}\hat{q}_{2} + S_1\hat{q}_{1}^2 + S_2\hat{q}_{2}^2.\label{Hamiltonian_a}
\end{eqnarray}
Introducing $\kappa\equiv {\epsilon_0 L^2}/{u_0^2\pi^2}$
the coefficients in~(\ref{Hamiltonian_a}) are
\begin{eqnarray}
&&D\equiv \frac{161 \pi^4 T_1}{4 L^2},\,\,F\equiv\frac{41\pi^4T_1}{2
  L^2},\,\,K_{1,2} = \kappa(V_1^2 \mp V_2^2),\nonumber \\
&& K_{12} = -\frac{16 \kappa}{9
  u_0}(V_1^2 - V_2^2),\,\,S_{1,2} = -{\kappa \pi^2 V_{1,2}^2}/{8 u_0}
\end{eqnarray}
For time dependent $V_{1,2}(t)$ the Hamiltonian (\ref{Hamiltonian_a}) describes
the excitation and evolution of two (nearly) degenerate
interacting flexural modes. Also other modes, not
included in ~(\ref{Hamiltonian_a}), will be excited by
the gates. This two-mode approximation is valid for weak intermode
interaction and when the other modes are off resonance with the modes
${\bf1}$ and ${\bf2}$.


For cat state generation we analyze the evolution of the system,
initially in the ground state, subject to a common bias pulse on the
two gates, i.e., $V_1=V_2=V_0\theta(t)$ where $\theta(t)$ is
the unit step function. Then (\ref{Hamiltonian_a}) reduces to
\begin{equation}
H=\hbar\sum_{j=1,2}\omega[\hat a_j^\dagger \hat a_j+
  \frac{\varepsilon}{4}(\hat a_j+\hat a_j^\dagger)^4]+\hbar\omega
\delta(t)(\hat a_2+\hat a_2^\dagger)+H_{12},
\label{eq:catham}
\end{equation} 
where $\omega^2\equiv\omega_{1,2}^2 + 2S_{1,2}/M$, $H_{12}=F\hat
q_1^2\hat q_2^2$, $\varepsilon={D\hbar}/{M^2\omega^3}$, and
$\delta(t)={\sqrt{2}\kappa V_0^2}/{\sqrt{\hbar M \omega^3}}$. The
$\hat a_{j}^{(\dagger)}$ are defined through $\hat{p}_{j}=
ip_0(\hat{a}_{j}^{\dag}- \hat{a}_{j})/\sqrt{2}$, and
$\hat{q}_{j}=x_0(\hat{a}_{j}^{\dag}+ \hat{a}_{j})/\sqrt{2}$ with
$x_0\equiv\sqrt{\hbar/M\omega}$ and $p_0\equiv \sqrt{M\hbar\omega}$.

We consider the situation when the system is cooled to $k_BT\ll
\hbar\omega$ and at time $t=0$ the flexural modes are in their
ground states $\left|0\right>$. Equal voltage pulses are applied to
both gates $\delta(t)=\delta_0\theta(t)$, inducing system's
evolution. As mode ${\bf1}$ will not be appreciably affected by the weak
coupling term $H_{12}$, the dynamics of the modes decouple.
Hence, mode ${\bf1}$ will remain close to its ground state for $t>0$. 
The remaining mode ${\bf2}$ describes a particle in a potential
$v(\xi)=\sqrt{2}\delta_0\xi+\xi^2/2+\epsilon\xi^4$ with $\xi=q_2/x_0$
and an equilibrium position $\xi_0$.  Introducing the displaced
oscillator operator $b_2 = a_2 -\xi_0/\sqrt{2}$, and applying the
rotating wave approximation (RWA), the Hamiltonian becomes $
H=\hbar\tilde{\omega}\hat{n}_b + 3\hbar\omega\varepsilon\hat{n}_b^2/2,
$
where
\be\label{eq:RWA} 
\hat{n}_b=\hat{b}^{\dag}_2\hat{b}_2,\:\:\:\:
\tilde{\omega}= \omega[1+3\varepsilon(1/2+2\xi_0^2)].
\ee
If $\epsilon\xi_0^2\ll1$, the situation is similar to
the one in~\cite{Yurke_Stoler}. The intial
state $\left|\psi(t=0)\right>=\left|0\right>_{a_2}$ resembles 
a coherent state in the $b_2$-basis, i.e,
$\left|0\right>_{a_2}\approx \left|-\xi_0/\sqrt{2}\right>_{b_2}$. If
the system evolves a time $T_1=\pi/(3\omega\varepsilon)$,
we expect to find the state
\begin{eqnarray}\nonumber
\vert\psi(T_1)\rangle\propto
e^{-i\pi/4}\vert-\xi_0 e^{-i\tilde{\omega}t}/\sqrt{2}\rangle_{b_2}+
e^{i\pi/4}\vert\xi_0 e^{-i\tilde{\omega}t}/\sqrt{2}\rangle_{b_2}\nonumber
\end{eqnarray}
which, to an overall phase, is in $a$-basis given by
\be\label{eq:cat_abasis}
\vert\psi(T_1)\rangle\propto
\vert\xi_0(1-e^{-i\tilde{\omega}t})/\sqrt{2}\rangle_{a_2}+
i\vert\xi_0(1+e^{-i\tilde{\omega}t})/\sqrt{2}\rangle_{a_2}. 
\ee
To verify this we simulated the dynamics using the full two-mode
Hamiltonian in ~(\ref{Hamiltonian_a}) and a Hilbert space of $40^2$
number states in the occupation basis. The parameters used were
$L=126$~nm, $u_0=60$~nm, $T_0=0.003$~N/m, and $V_0=0.23$~V. This
corresponds to $\omega/(2\pi)\approx 560$~MHz,
$\varepsilon=7\cdot10^{-4}$ and $\delta_0=1.3$. The position shift
becomes $\xi_0\approx-1.8\approx -\sqrt{2}\delta_0$.

\begin{figure}[t]
\begin{center}
\epsfig{file=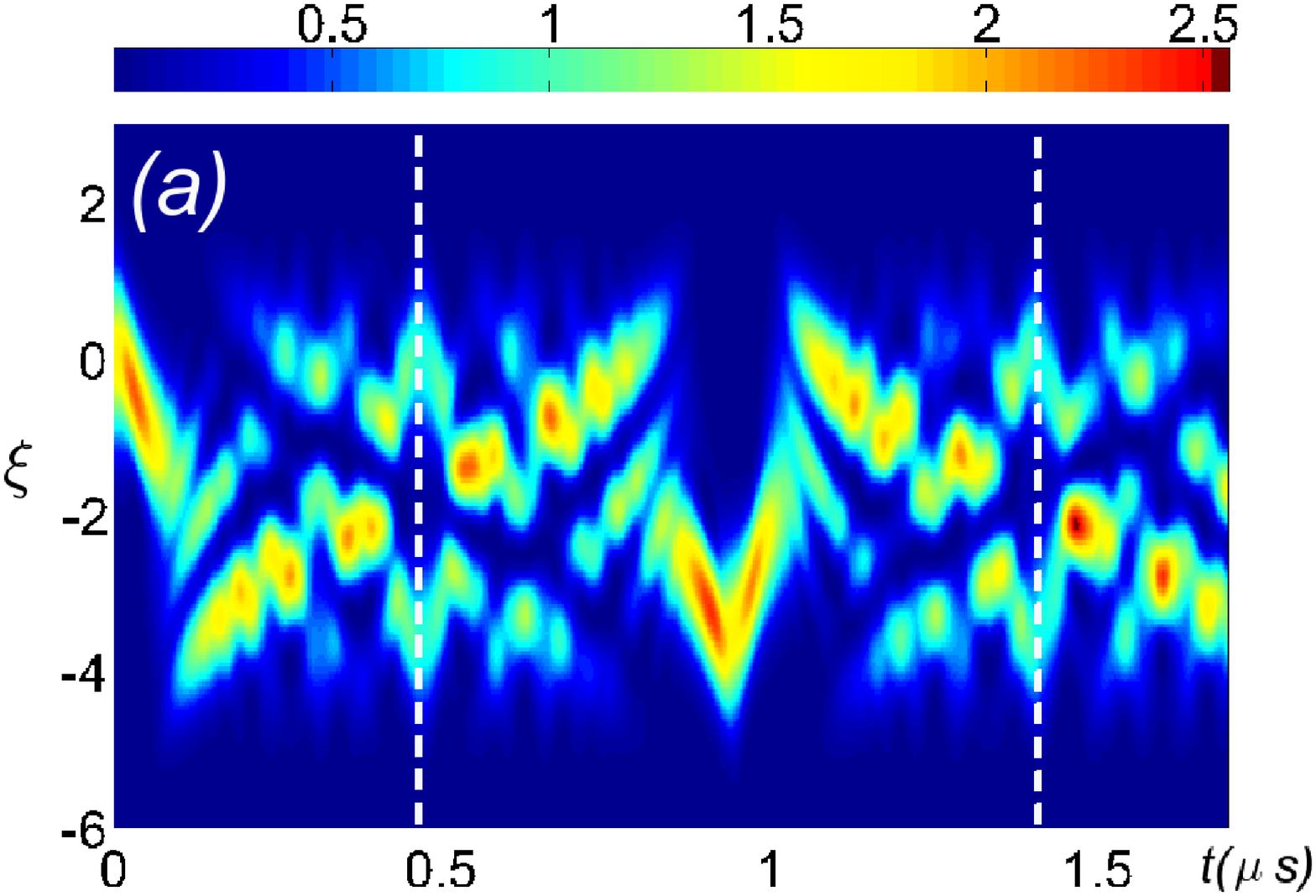, width=\linewidth}\\
\epsfig{file=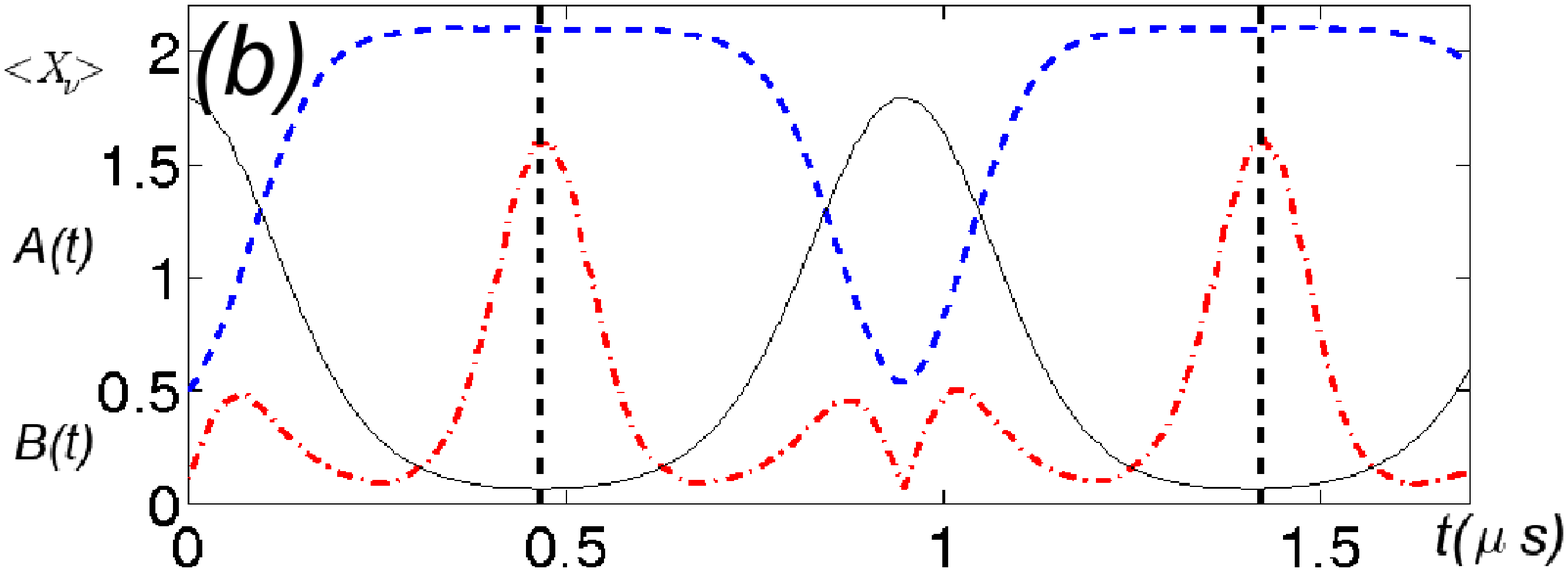, width=\linewidth}\\
\end{center}
\caption{(Color online) (a) False color plots of snapshots of time
  evolution of the position probability distribution for one flexural
  mode. The snapshots are taken at the turning points of the
  corresponding classical trajectory of the system. The positions
  (y-axis) are scaled to the quantum zero point fluctuations $x_0$.
  with vertical, white dashed lines. (b) Corresponding time evolution
  of the envelopes of the quadrature $\left< X_{\nu}\right>$
  (continous line) and its associated quantum fluctuations
  $\left<\Delta X_{\nu}^2\right>$; $A(t)$ (dashed line),
  $B(t)$ (dashdotted line). The appearance
  of a cat state is signalled by a decrease in $\left< X_{\nu}\right>$
  along with an increased contribtion from quantum fluctuations
  $\left<\Delta X_{\nu}^2\right>$ in the noise of
  $\left<X_{\nu}\right>$.\ Vertical dashed lines indicate the
  agreement with Fig.\ 2a.
 \label{fig:cats}}
\end{figure}

In Fig.~\ref{fig:cats}a the probability density of finding mode
${\bf2}$ at a position $\xi$ is shown as function of time. Only
instants where $t=2\pi n/\tilde{\omega}$ are sampled hence fast
oscillations with frequency $\omega$ are not visible. The initial
state $\left|0\right>_{a_2}= \left|-\xi_0/\sqrt{2}\right>_{b_2}$
evolves into a first cat-like state at time
$T_1\approx0.42$~$\mu$s~$\approx\pi/(3\omega\varepsilon)$. This and
the following cat-like state emerging at $T_2\approx1.27$~$\mu$s~
$\approx\pi/(\omega\varepsilon)$ are marked with vertical
dashed lines. The simulations also verified that mode ${\bf1}$
remains close to its ground state.

To read out the state the two gates may form part of a capacitor in an
LC-circuit with $\omega_{\rm LC}\gg \omega$ in the resolved sideband
limit. This allows for side band cooling and measurement of the
quadrature operator~\cite{Hertzberg_2009}
$X_\nu=[e^{i\nu}a^{\dagger}_2+e^{-i\nu}a_2 ]/\sqrt{2}$. In
Fig.~\ref{fig:cats}b we show the envelopes of the average $\left<
X_\nu\right>$ along with the quantum fluctuations $\left<\Delta
X_\nu^2\right>$ as functions of time. The fluctuations have the form
$\left<\Delta X_\nu^2\right>\sim A(t)+B(t)\cos(2\omega t)$, where $A$
and $B$ vary slowly in time. A signature of the built up cat state is
the reduction of $\left< X_\nu\right>$ with an increase in
$\left<\Delta X_\nu^2\right>$.

For the cat state to emerge, dissipation must be weak. Even at zero
temperature damping will cause decoherence. As shown
in~\cite{Holmes_1986, Kartner_1993} this occurs on a time scale
$t_c\sim Q/\omega\xi_0^2$ where $Q$ is the resonator quality factor.
Requiring $t_c\gg T_1$ yields $Q\gg \xi_0^2/\varepsilon$. For our
protoype system, this inequality is fulfilled if
$Q>10^4$. Recently Q-factors up to $Q\sim 10^5$ were reported in
graphene resonators~\cite{Bachtold_2011}.

The cat state above is a product state between modes ${\bf1}$ and
${\bf2}$ in $a$-basis. We now demonstrate a cat-like state involving
both modes that is not a product state in this basis. Preparing the
system in the ground state $\left|0,0\right>_a$ and switching on only
one gate at time $t=0$, i.e., $V_2=0$, $V_1=\tilde{V}$, the degeneracy
is lifted. Just as the operators $a^{(\dag)}_j$ and $b^{(\dag)}_j$
respectively diagonalize the linear part of (\ref{Hamiltonian_a}) when
$V_{1,2}=0$ and $V_1=V_2\neq 0$, the normal modes when $V_1\neq V_2
=0$ are found by diagonalizing the linear
part of (\ref{Hamiltonian_a}) by means of the transformation
\begin{eqnarray}
&&d_1 = \cos(\theta) a_1 -\sin(\theta) a_2\nonumber,\\ 
&&d_2=\sin(\theta) a_1 +\cos(\theta) a_2+\eta_0.\nonumber
\end{eqnarray} 
Here $\eta_0=\kappa\tilde{V}^2/\sqrt{\hbar M\omega^3}$ and we have
neglected terms of order $|(\Omega_1-\Omega_2)/\omega|\ll 1$, where
$\Omega_{1,2}^2=\omega^2 \mp K_{12}/M$ are the new eigenmode
frequencies. The Hamiltonian (\ref{Hamiltonian_a}) now transforms
to
\be
\label{eq:Hamiltonian_d}
\tilde{H} = \hbar\sum_{j=1,2}\Omega_jd_j^{\dag}d_j
+\tilde{H}_{NL}, 
\ee 
where $\tilde{H}_{NL}$ is quartic in $d$-operators. As parameters like
length and tension are never exactly equal in both $x$ and
$y$-directions, the two modes ${\bf1,2}$ are only approximately
degenerate, i.e. $\omega_1\neq \omega_2$. This leads to the
requirement $|(\Omega_1-\Omega_2)/(\omega_1-\omega_2)|\gg 1$ so that
$\theta\simeq\pi/4$, and
\begin{eqnarray}\label{eq:H_NL}
\tilde{H}_{NL}&=& \frac{\tilde{\varepsilon}}{4}(d_1+d_1^{\dag})^4+
\frac{\tilde{\varepsilon}}{4}(d_2 + d_2^{\dag}- 2\eta_0)^4
+\\\nonumber 
&&\lambda(d_1+d_1^{\dag})^2(d_2 +d_2^{\dag}- 2\eta_0)^2,\nonumber
\end{eqnarray}
\be\nonumber
\tilde{\varepsilon}={\hbar(2D-F)}/{4M^2\omega^3},\:\:\:\:
\lambda={\hbar(6D-F)}/{2M^2\omega^3}.\\\nonumber 
\ee 
The initial state $\left|0,0\right>_a$ is in $d$-basis
$\left|0,\eta_0\right>_{d}$.
  After evolution with
(\ref{eq:Hamiltonian_d}) this state will, analogous to the situation
studied above, enter a superposition at time $\tilde{T}_1\approx
\pi/(3\omega\tilde{\varepsilon})\approx 0.65 \mu s$
\begin{eqnarray}\label{eq:superpos_oscd2}
\vert\psi(\tilde{T}_1)\rangle& \propto&
\left|0,\eta_0\right>_d +
i\left|0,-\eta_0\right>_d +
\alpha\left|\chi\right>.
\end{eqnarray}
Here $\alpha\left|\chi\right> = \alpha \sum_{n=1}
\left|n\right>_{d_1}\left|\Psi_n\right>_{d_2} $ is due to
the quartic coupling term in (\ref{eq:H_NL}). The state component
$\left|0,\eta_0\right>_d + i\left|0,-\eta_0\right>_d$ corresponds to a
cat-like non-product state $\left|0,0\right>_a +
i\left|-\sqrt{2}\eta_0,-\sqrt{2}\eta_0\right>_a$ in $a$-basis. 

To verify the creation of this state we numerically analyze the
evolution of $\vert 0,0 \rangle_{a}$. Fig.\ 3a shows the Wigner
distribution $W(\xi_{1},\pi_{1})=(2\pi)^{-1}\int d\xi~ e^{-i\pi_{1}
  \xi} \langle \xi_{1} +\xi/2\vert \hat{\rho}_{1}\vert
\xi_{1}-\xi/2\rangle$ of the reduced density matrix $\hat{\rho}_{1}$
of mode ${\bf1}$ in $a$-basis at $\tilde{T}_1$. The distribution is
plotted as function of the dimensionless position $\xi_{1}$ and
momentum $\pi_{1}$ and has a bimodal structure. The distribution for
mode ${\bf2}$ is identical, $W(\xi_2,\pi_2) = W(\xi_1,\pi_1)$.

To remove $\alpha\vert\chi\rangle$ from (\ref{eq:superpos_oscd2}) we
introduce the projection operators
$\hat{P}_n=\left(\left|n\right>\left<n\right|\right)_{d_1}\otimes
\hat{I}_{d_2}$ and study (Fig.~3b) the Wigner distribution of the
projection $\hat{P}_0\vert\psi(\tilde{T}_1)\rangle$. One can here
clearly recognize the distribution corresponding to the state
$\left|0,0\right>_a +
i\left|-\sqrt{2}\eta_0,-\sqrt{2}\eta_0\right>_a$, which is displayed
in Fig.\ 3c for reference.  To ensure that the
$\alpha\vert\chi\rangle$ component is not of major significance, the
time evolution of the projections $\langle\hat{P}_0(t)\rangle$,
$\langle\hat{P}_2(t)\rangle$ and
$\langle\hat{I}-\hat{P}_0(t)-\hat{P}_2(t)\rangle$ are shown in
Fig.\ 3d.

\begin{figure}[htpb]
\begin{center}
\begin{eqnarray*}
\includegraphics[width=0.5\linewidth]{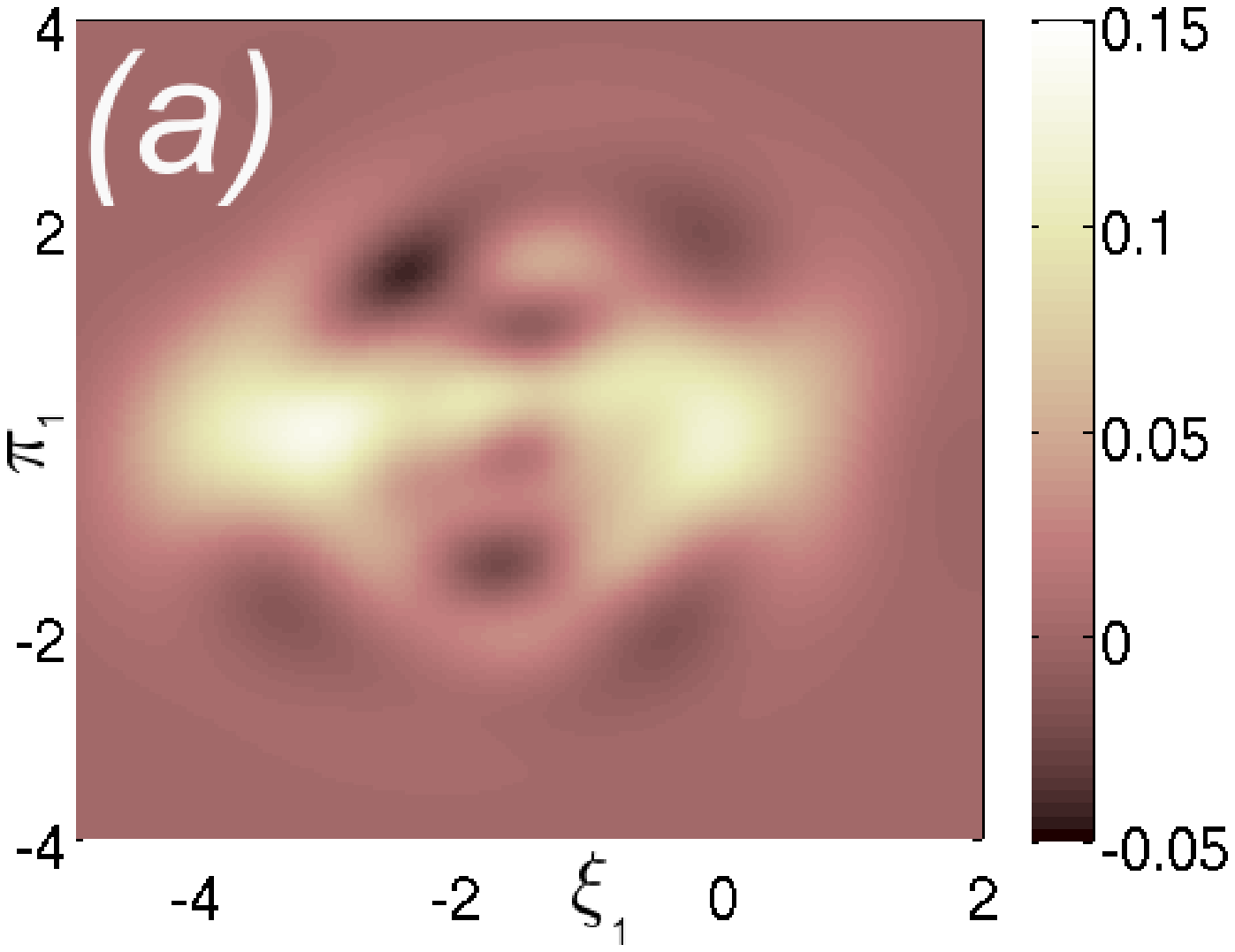} 
\includegraphics[width=0.5\linewidth]{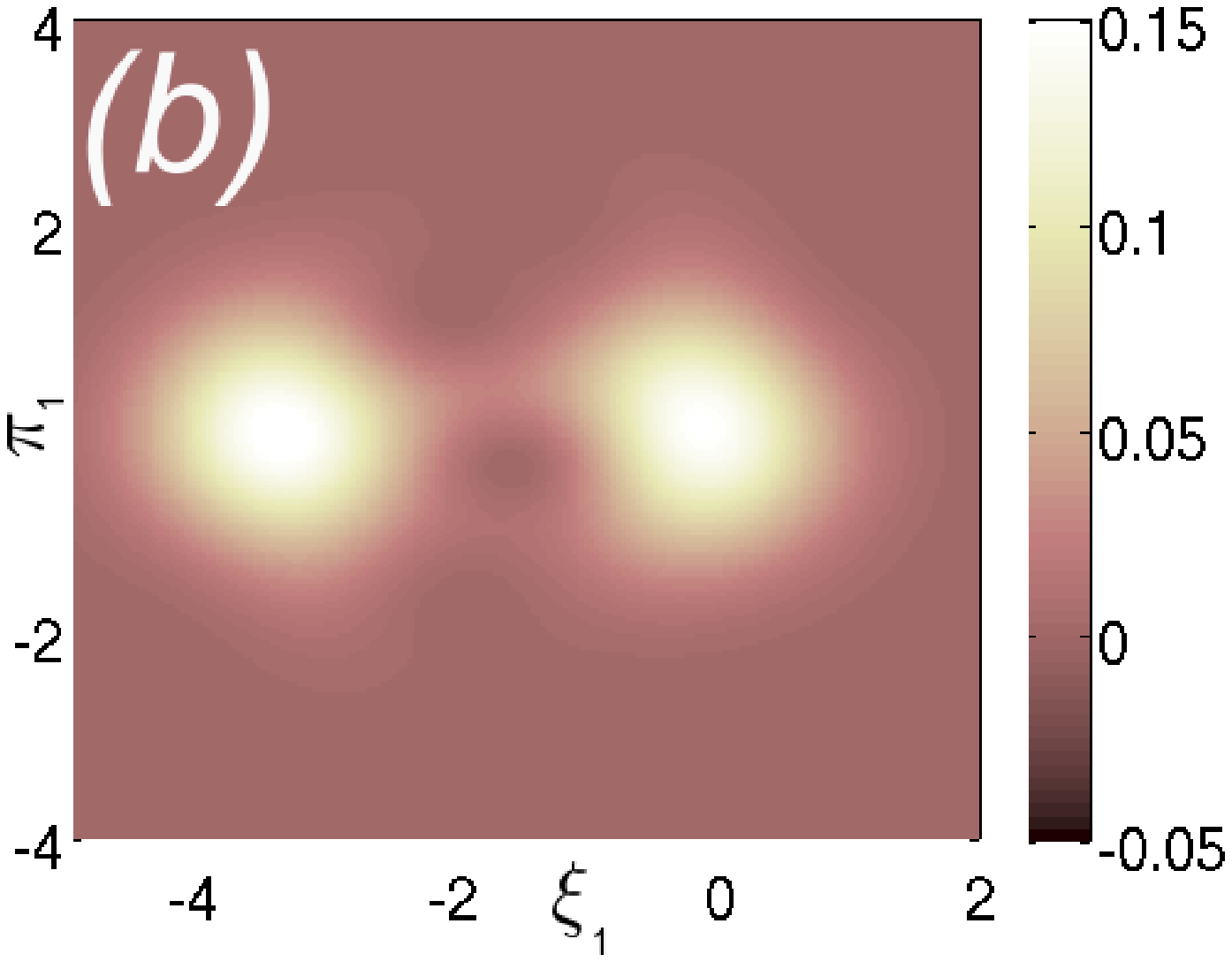} \\
\includegraphics[width=0.5\linewidth]{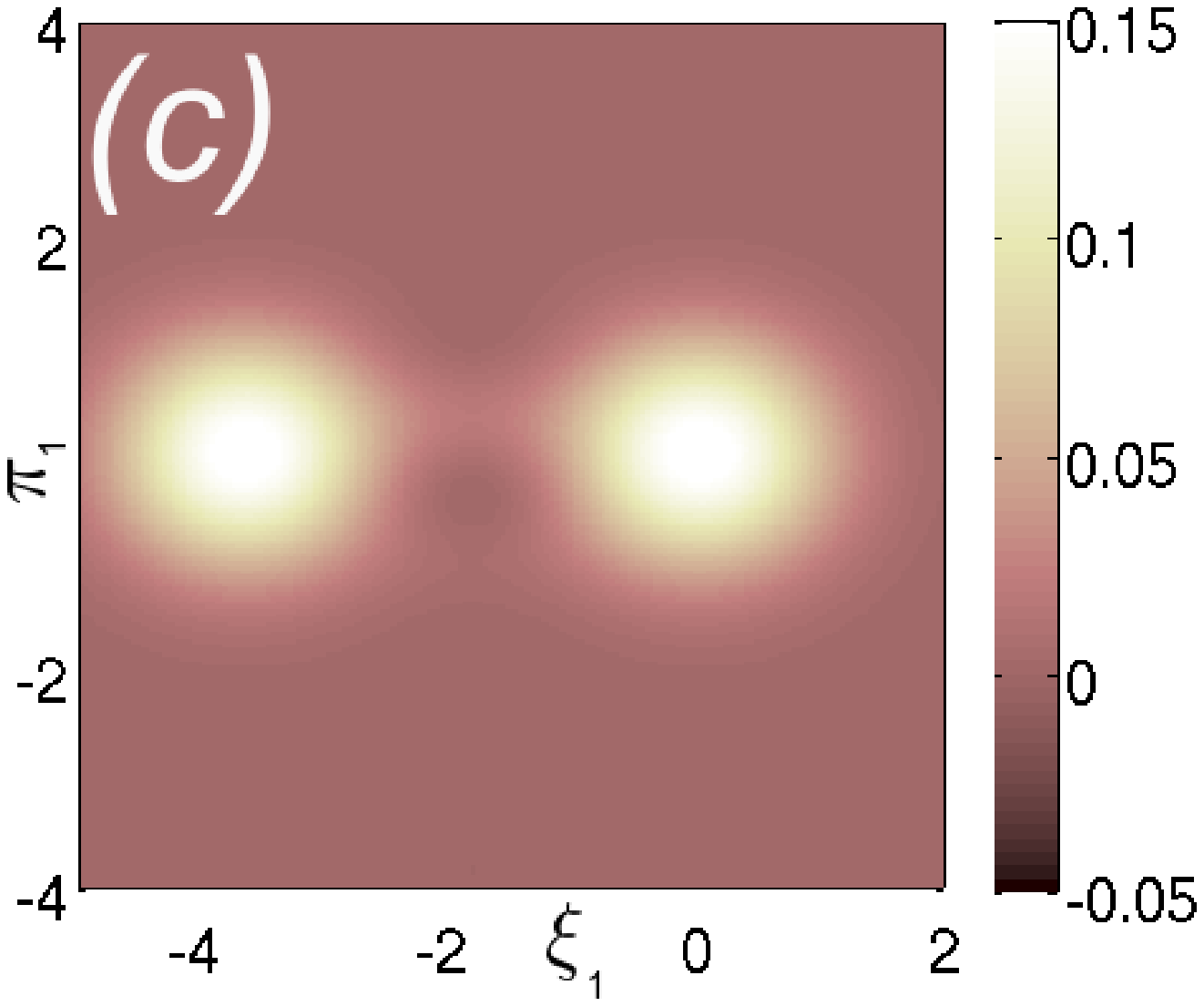}
\includegraphics[width=0.5\linewidth]{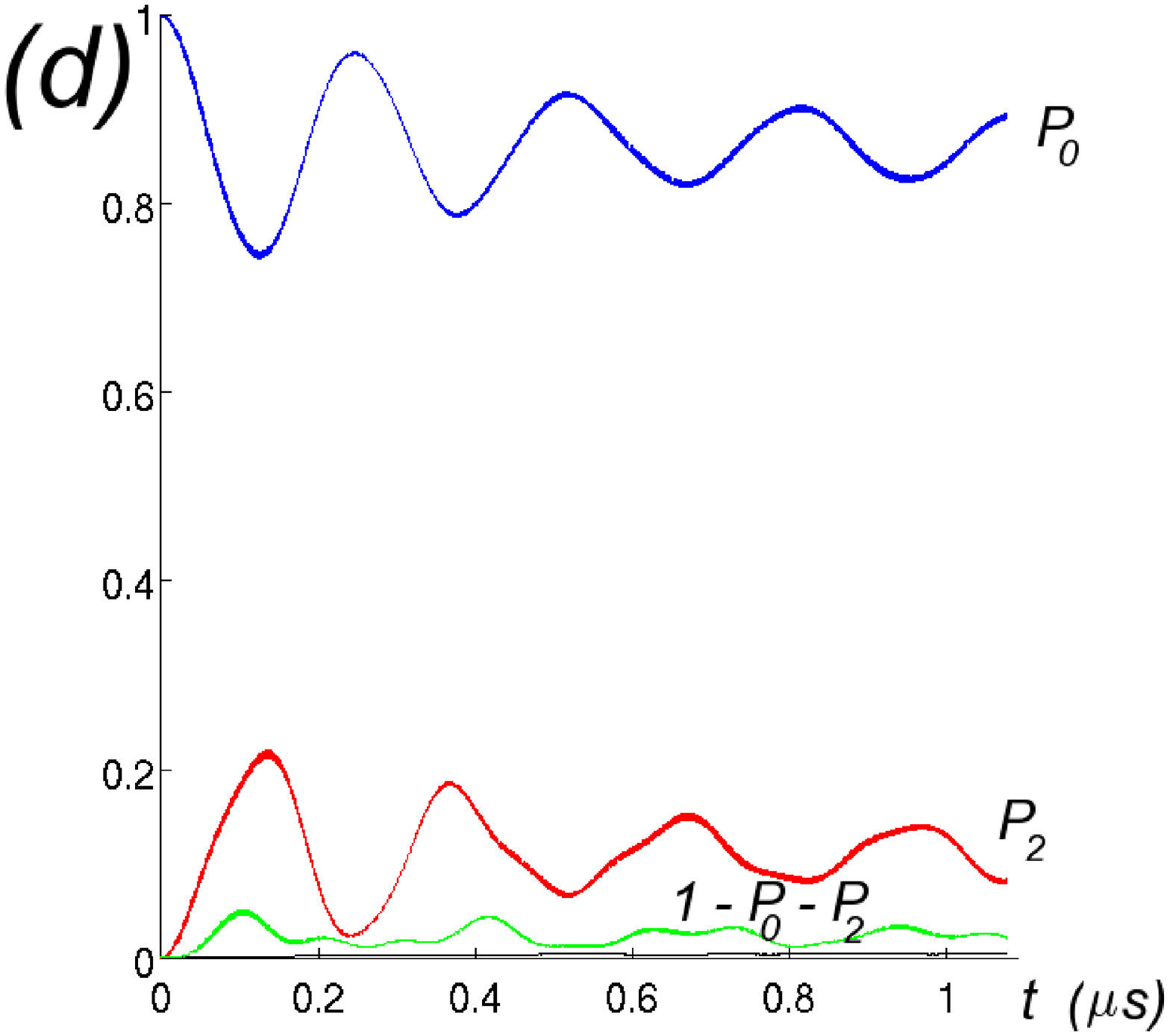} 
\end{eqnarray*}
\end{center}
\caption{(Color online) (a) False color plot of reduced Wigner distribution of the
  time evolved initial state $\vert 0,0\rangle_{a}$, sampled at
  $\tilde{T}_1$ as function of the dimensionless position $\xi_{1}$
  and momentum $\pi_{1}$. The reduced Wigner distributions are
  identical for both modes $W(\xi_1,\pi_1)=W(\xi_2,\pi_2)$. A bimodal
  structure is seen. (b) False color plot of Wigner distribution of
  $\hat{P}_0\hat{\rho}_{1}(\xi_{1},\pi_{1},\tilde{T}_1)\hat{P}_0$,
  clearly demonstrating bimodality. (c) False color plot of
  Wigner distribution of $\left|0,0\right>_a +
  i\left|-\sqrt{2}\eta_0,-\sqrt{2}\eta_0\right>_a$. The similarity to
  the projection in (b) is evident. (d) Time evolution of projections
  $\langle\hat{P}_0(t)\rangle$, $\langle\hat{P}_2(t)\rangle$ and
  $\langle\hat{I}-\hat{P}_0(t)-\hat{P}_2(t)\rangle$. The most
  significant contribution comes from
  $\langle\hat{P}_0(t)\rangle$.}\label{fig:wigner}
\end{figure}

The expression for 
$\vert\psi(\tilde{T}_1)\rangle$ in (\ref{eq:superpos_oscd2}) is in
RWA. In the Schr{\"o}dinger picture the state has a fast
oscillating component and is at $t=\tilde{T}_1$ 
\begin{eqnarray}\nonumber
&&\vert\psi(t)\rangle_S=\vert 0,\:\eta_0 e^{-i\tilde{\Omega}_2
  t}\rangle_d + i\vert0,-\eta_0 e^{-i\tilde{\Omega}_2
  t}\rangle_d + \alpha\vert\chi\rangle \\\nonumber
&&= \left|\frac{\eta_0}{\sqrt{2}} (e^{-i\tilde{\Omega}_2t}-1),
\frac{\eta_0}{\sqrt{2}}(e^{-i\tilde{\Omega}_2t}-1)\right>_a\\\nonumber
&&+i \left| -\frac{\eta_0}{\sqrt{2}}(e^{-i\tilde{\Omega}_2t}+1),-\frac{\eta_0}{\sqrt{2}}(e^{-i\tilde{\Omega}_2t}+1)\right>_a+ \alpha\vert \chi \rangle .\nonumber
\end{eqnarray}
Here $\tilde{\Omega}_2$ is the eigenfrequency renormalized due to the
nonlinearities in (\ref{eq:H_NL}) [cf. (\ref{eq:RWA})]. 

Similar behavior is seen for evolution from the initial state
$\left|\eta_0,0\right>_d =\left|0,-\sqrt{2}\eta_0\right>_a $.
Bimodality is then observable at $\tilde{T}_2 \approx 0.75 \mu s$. We
attribute the difference $\tilde{T}_2-\tilde{T}_1$ to the position
shift $2\eta_0$ in (\ref{eq:H_NL}).

Finally we demonstrate cat-like non-product states in both $a-$ and
$d-$bases. Assume the cat state (\ref{eq:cat_abasis}) was generated by
the two-gate configuration at $T_1$. One gate is then switched off
when $\tilde{\omega} T_1= 2\pi m$, $m$=integer. This kind of
time-domain control has been shown possible
in~\cite{Freeman_2008}. The cat state is then a superposition of
coherent states in $d$-basis $\left|\psi(T_1)\right>\propto
\left|0,~\eta_0\right>_d +i\left|-\xi_0,~\xi_0+\eta_0\right>_d$. If
$\tilde{V}=\sqrt{2}V_0$, then $\eta_0\approx -\xi_0$, and the state is
\begin{equation}
\left|\psi(T_1)\right> \approx \left|0,\eta_0\right>_d+
i\left|\eta_0,0\right>_d.
\label{eq:catstate_dbasis}
\end{equation}
As in previous cases one would expect that after an
evolution with (\ref{eq:Hamiltonian_d}) both modes would enter a superposition
\begin{eqnarray}\nonumber
&&\left|0,\eta_0\right>_d
+i\left|0,-\eta_0\right>_d + i\left(\left|\eta_0,0\right>_d +i\left|-\eta_0,0\right>_d\right)+\beta\vert \zeta
\rangle, \nonumber
\label{eq:superpos_2modes}
\end{eqnarray}
where $\beta\vert \zeta \rangle$ again is a small remainder due to the
quartic coupling. Numerically we observe cat-like states in both modes
in time interval $0.72 \mu$s $< \tilde{T}_1 < 0.82 \mu$s. These are
cat-like superposition states which are not product states in either
$a$- or $d$-basis.

We have shown that due to the intrinsic non-linearities in graphene,
generation of cat states and multimode cat states is possible by local
back-gate manipulation. The nonlinearities are strong enough to avoid
the necessity of coupling the membrane to an auxilliary
system. Together with recent advancement in graphene device
fabrication and improvement of cooling schemes this opens
for further fundamental studies of macroscopic quantum phenomena.

The research leading to these results has received funding [AV, AI]
from the EU 7$^{th}$ framework programme (FP7/2007-2013) QNEMS (grant
agreement no: 233992) and the Swedish Research Council [JMK].
We also thank J. Atalaya and A. Croy for stimulating
discussions.


\begin{thebibliography}{99}
\bibitem{Monroe_1996} C. Monroe et.al, Science {\bf 272}, 1131 (1996).
\bibitem{Friedman_2000} J. R. Friedman et.al, Nature {\bf 406}, 43 (2000).
\bibitem{Haroche_2008} S. Deléglise et.al, Nature {\bf 455}, 510
  (2008).
\bibitem{Schwab_2005} K. C. Schwab, and M. L. Roukes, Phys. Today {\bf
  58},36 (2005).
\bibitem{Cleland_2010} A. D. O’Connell et.al,Nature {\bf 454}, 697 (2010).
\bibitem{Lehnert_2011} J. D. Teufel et.al, Nature {\bf 475}, 359 (2011).
\bibitem{Cleland_2009} M. Hofheinz et.al, Nature {\bf 459}, 546
  (2009).
\bibitem{Gardiner_book} Quantum Noise, C. W Gardiner, P. Zoller,
  Springer Verlag (2000). 
\bibitem{Yurke_Stoler} B. Yurke, and D. Stoler, Phys. Rev. Lett. {\bf
  57}, 13 (1986).
\bibitem{Cleland_book} A. N. Cleland, {\it Foundations of Nanomechanics} (Springer, New York, 2003).
\bibitem{Holmes_1986} G. J. Milburn, and C. A. Holmes,
  Phys. Rev. Lett. {\bf 56}, 2237 (1986).
\bibitem{Kartner_1993} F. X. Kartner, and A. Schenzle, Phys. Rev. A
  {\bf 48},1009 (1993).
\bibitem{Jacobs_2007} K. Jacobs, Phys. Rev. Lett. {\bf 99}, 117203
  (2007).
\bibitem{Milburn_2009} F. L. Semiao, K. Furuyu, and G. J. Milburn,
  Phys. Rev. A {\bf 79}, 063811 (2009).
\bibitem{Atalaya_2008} J. Atalaya, A. Isacsson, and J. M. Kinaret,
  Nano Lett. {\bf 8}, 4196 (2008).
\bibitem{Bachtold_2011} A. Eichler et.al, Nat. Nanotechn. {\bf 6}, 339 (2011).
\bibitem{Tombesi_2002} S. Mancini, V. Giovanetti, D. Vitali, and
  P. Tombesi, Phys. Rev. Lett. {\bf 88}, 12 (2002).
\bibitem{Plenio_2008} M. J. Hartmann, and M. B. Plenio,
  Phys. Rev. Lett. {\bf 101}, 200503 (2008).
\bibitem{Marquardt_2010} M. Ludwig, K. Hammerer, and F. Marquardt,
  Phys. Rev. A {\bf 82}, 012333 (2010).
\bibitem{Zhang_2011} L. Zhou, Y. Han, J. Jing, and W. Zhang,
  Phys. Rev. A {\bf 83}, 052117 (2011).
\bibitem{Girvin_2011} K. Borkje, A. Nunnenkamp, and S. M. Girvin,
  Phys. Rev. Lett. {\bf 107}, 123601 (2011). 
\bibitem{Sanders_1992} B. C. Sanders, Phys. Rev. A {\bf 45}, 6811
  (1992).
\bibitem{Daniel_book} Vibrations of Shells and Plates, W. Soedel,
  Marcel Dekker Inc (2004).
\bibitem{Yakobsson_2001} K. N. Kudin, G. E. Scuseria, and
  B. I. Yakobsson, Phys. Rev. B {\bf 64}, 235406 (2001).
\bibitem{Hertzberg_2009} J. B. Hertzberg et.al, Nat. Phys. {\bf 6}, 213
  (2010).
\bibitem{Freeman_2008}N. Liu et.al, Nat. Nanotechnology {\bf 3}, 715 (2008). 
\end{thebibliography}
\end{document}